\documentclass[%
 reprint,%
 amssymb, amsmath,%
 aip,cha%
]{revtex4-1}
 \usepackage{graphicx}%
\usepackage{docs}%
\usepackage{bm}%
\usepackage{amsmath}%
\usepackage[colorlinks=true,linkcolor=blue]{hyperref}%

\expandafter\ifx\csname package@font\endcsname\relax\else
 \expandafter\expandafter
 \expandafter\usepackage
 \expandafter\expandafter
 \expandafter{\csname package@font\endcsname}%
\fi
\graphicspath{{../}}

\begin{document}

\title{Imaging Strain and Electric Fields in NV Ensembles using Stark Shift Measurements}%

\author{S. Sharma}%
\email{ssharm18@illinois.edu}
\affiliation{University of Illinois, 1110 W Green St, Urbana, IL 61801, USA}%
\author{C. Hovde}%
\affiliation{Southwest Sciences Ohio Operations, 6837 Main St, Newtown, OH 45244, USA}%
\author{D.H. Beck}%
\affiliation{University of Illinois, 1110 W Green St, Urbana, IL 61801, USA}%

\author{ F. Alghannam}%
\affiliation{Texas A\&M University, 400 Bizzell St, College Station, TX 77843, USA}%

\date{December 2017}%

\begin{abstract} % abstract
We report measurements of optically detected magnetic resonance spectra of ensembles of negatively charged nitrogen-vacancy (NV) centers in diamonds in the presence of strain and DC external electric fields. The Stark shift of the spectral lines is stronger when the axial magnetic field along the NV center's quantization axis is minimized.  The shift is also enhanced at avoided crossings between the hyperfine levels at an axial field of $B = \pm$ 77 $\mu T$.   Since the intrinsic strain in the diamond also induces a Stark shift, we are able to calculate the magnitude and direction of the strain within the crystal. We also use the Stark effect to map the electric field in the diamond volume between patterned electrodes.
\end{abstract}

\maketitle

\begin{figure}
{\includegraphics[scale = 0.255]{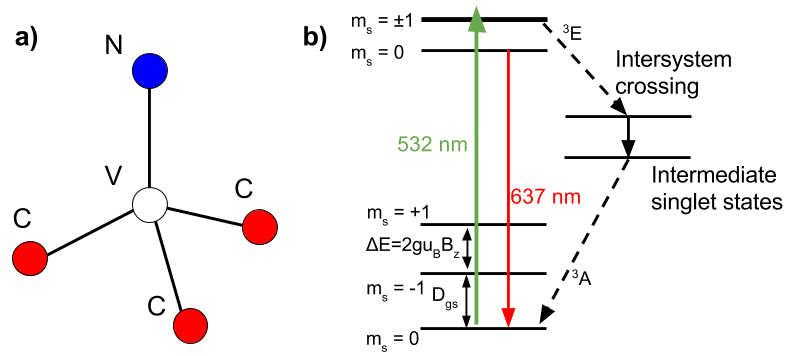}}
\caption{a) Structure of an NV center in the diamond lattice. The quantization axis of the NV center is along the line connecting the nitrogen
atom and the vacancy. b) Relevant spectroscopy of NV centers, including ground state fine structure (not to scale). Only a single arm of the excited state doublet is shown. Hyperfine structure is also omitted for simplicity.}
\label{fig:EnergyLevels}
\end{figure}

\section{Introduction}

Nitrogen-vacancy centers are formed in diamonds when a nitrogen atom substitutes for one of the carbon atoms in the lattice and it pairs with a vacancy. A center can acquire an electron to form  NV$^-$ , which we hearafter indicate as NV. The defect center can also be electrically neutral (NV$^0$) or carry a positive charge as the optically inactive(NV$^+$).\cite{NVPlus}

NV centers have relatively long-lived spin coherence features that can be initialized and detected with visible light.\cite{dohertyReview}  Recent studies of single NV centers are motivated by applications including quantum computing \cite{Childress} and nanoscale sensing of magnetic \cite{bala} and electric\cite{Dolde} fields. 

The electronic ground state $^3A_2$ is an orbital singlet state and consists of a $m_s=0$  level  separated from nearly degenerate $m_s=\pm 1$ levels by $D_{gs}$ $\approx$ 2.87 GHz (Fig. \ref{fig:EnergyLevels}).  The NV center can absorb off-resonant 532 nm light, after which it fluoresces from the $^3$E excited state. The zero-phonon line(ZPL) is 637 nm, but significant fluorescence is observed in wavelengths up to 800 nm. A competing non-radiative pathway via intermediate singlet states also exists for the excited state and is stronger for the excited state m$_s$=$\pm$1 levels. Because of this non-radiative pathway, the overall fluorescence of the excited state depends on the excited state $m_s = \pm 1$ population which in turn depends on the ground state $m_s = \pm 1$ population.  This allows the spin of the ground state of the NV center to be determined by monitoring the brightness of the emitted fluorescence.\cite{dohertyReview} 

 Either CW or pulsed microwave techniques can be used to examine magnetic dipole transitions from the  $m_s=0$ to $m_s=\pm$1 levels in the ground state with high spectral resolution.\cite{dohertyReview} This technique of resolving the ground state electronic structure is called optically detected magnetic resonance (ODMR) and we use it in this work to measure DC Stark shifts. 
%The same high-resolution information can be obtained with nonlinear optical spectroscopy, such as coherent population trapping\cite{Santori} or electromagnetically-induced transparency\cite{AcostaEIT}.

%Van Oort and Glasbeek\cite{Vanoort} studied the Stark effect in ensembles of NV centers using Hahn spin-echo techniques. An electric field was applied during part of the pulse sequence, resulting in a sensitive measurement of the phase shift due to the electric field. Their work showed a linear Stark shift, indicating that the NV center lacked inversion symmetry.
%Electric fields can also be produced inside the diamond by photoionizing NV centers to NV$^0$, or by photoionizing other impurities. Photo-ionization is thought to explain the sudden changes in the wavelength of the $^3$E - $^3$A transition, a phenomenon known as spectral diffusion\cite{Tamarat}. 
%Recent work has shown that applied electric fields can be used to stabilize the absorption/emission wavelength of single NV centers\cite{AcostaTamarat}. Stability helps in coupling multiple NV centers as qubits.

Recent work\cite{Dolde} has shown how a single NV can be used for measuring the transverse component of the DC electric field and how NV ensembles can be used to measure electric fields applied across a 300 um wide diamond. \cite{Braje} 
%Taking strain induced Stark shifts into account, such measurements from NVs oriented along all four crystallographic axes would allow the full reconstruction of the electric field vector. 

We are interested in developing the capability to measure electric and magnetic fields near the interaction region of experiments to measure the electric dipole moment of the neutron.\cite{nEDM} Such experiments seek to push the boundaries of the Standard Model.
% and require fields that are stable and homogeneous over periods of about one year. 
%In some proposed experiments, the sensing element must operate at cryogenic temperatures, with optical access only through optical fibers. 
It is critical that the probe makes only small perturbations to the fields that it is measuring. In this regard, NV diamonds show promise, as the diamond is electrically and magnetically inert. While this work uses microwaves, previous work has shown that magnetic\cite{AcostaEIT}$^,$\cite{Sharma} and electric field\cite{Tamarat} effects can be detected with all-optical NV diamond probes. In this paper, we present measurements of strain and extenral electric fields using ODMR in NV ensembles.

To compare our results with theory, we model the  ground state Hamiltonian, $H_{gs}$ as\cite{Doherty}
\begin{equation}
\hat{H}_{gs} = \hat{H}_{hfs} + \hat{H}_{es}  \, .
\end{equation}
Here $ \hat{H}_{hfs}$ is the zero-field Hamiltonian which includes the hyperfine interactions with the $^{14}$N (spin = 1) nucleus. $ \hat{H}_{es}$ is the electronic spin Hamiltonian which includes the strain, magnetic and electric field interactions. 

The hyperfine interaction Hamiltonian is given by
\begin{equation}
\hat{H}_{hfs} = \dfrac{1}{\hbar^2} [D_{gs} S_z^2 + A_\parallel S_z I_z +A_\perp (S_xI_x + S_yI_y) + PI_z^2] \, .
\end{equation}
$D_{gs} \approx 2.87$   GHz  is the ground state zero-field splitting which arises from spin-spin interaction of the electrons and splits the $m_s = 0$ and $m_s = \pm 1$ sublevels as shown in Fig. \ref{fig:EnergyLevels}.  $A$ and $P$ are the magnetic hyperfine parameter and the nuclear quadrupole  parameter respectively. The solution to Eq.1 includes 9 eigenstates, 3 associated with $m_s$ = 0 and 6 associated with $m_s = \pm$ 1 levels.  Pairs of these six undergo avoided crossings, or level repulsions, at $B_{axial}$ = 0, $\pm$ 77 uT,  as described below. 

The electronic spin component of the ground state energy is given by
\begin{multline}
\hat{H}_{es} = \dfrac{1}{\hbar^2} d_\parallel \Pi_z S_z^2 + \dfrac{\mu_B}{\hbar} \vec{S} . \bar{g} .\vec{B} - \dfrac{1}{\hbar^2} d_\perp \Pi_y (S_x^2  - S_y^2) \\ - \dfrac{1}{\hbar^2} d_\perp \Pi_x (S_xS_y  + S_yS_x)
\end{multline}
Here, $\vec{ \Pi} = \vec{E} + \vec{\sigma}$ is the total effective electric field which also includes contributions from the strain, $\vec{\sigma}$. The ground state electric dipole moment, $d$, has a much larger transverse component $d_{\perp} \approx 17  $ Hz cm/V compared to the longitudinal component,   $d_{\parallel} \approx 0.3  $ Hz cm/V. The effective g-factor tensor, $\bar{g}$ has a longitudinal, $g_\parallel$ and a transverse, $g_\perp$ component as detailed in Ref. 6 % \cite{Doherty}. Longitudinal electric fields effectively produce a shift in  $D_{gs}$ while transverse electric fields make the $m_s= \pm1$ levels repel. 

\begin{figure}
{\includegraphics[scale = 0.15]{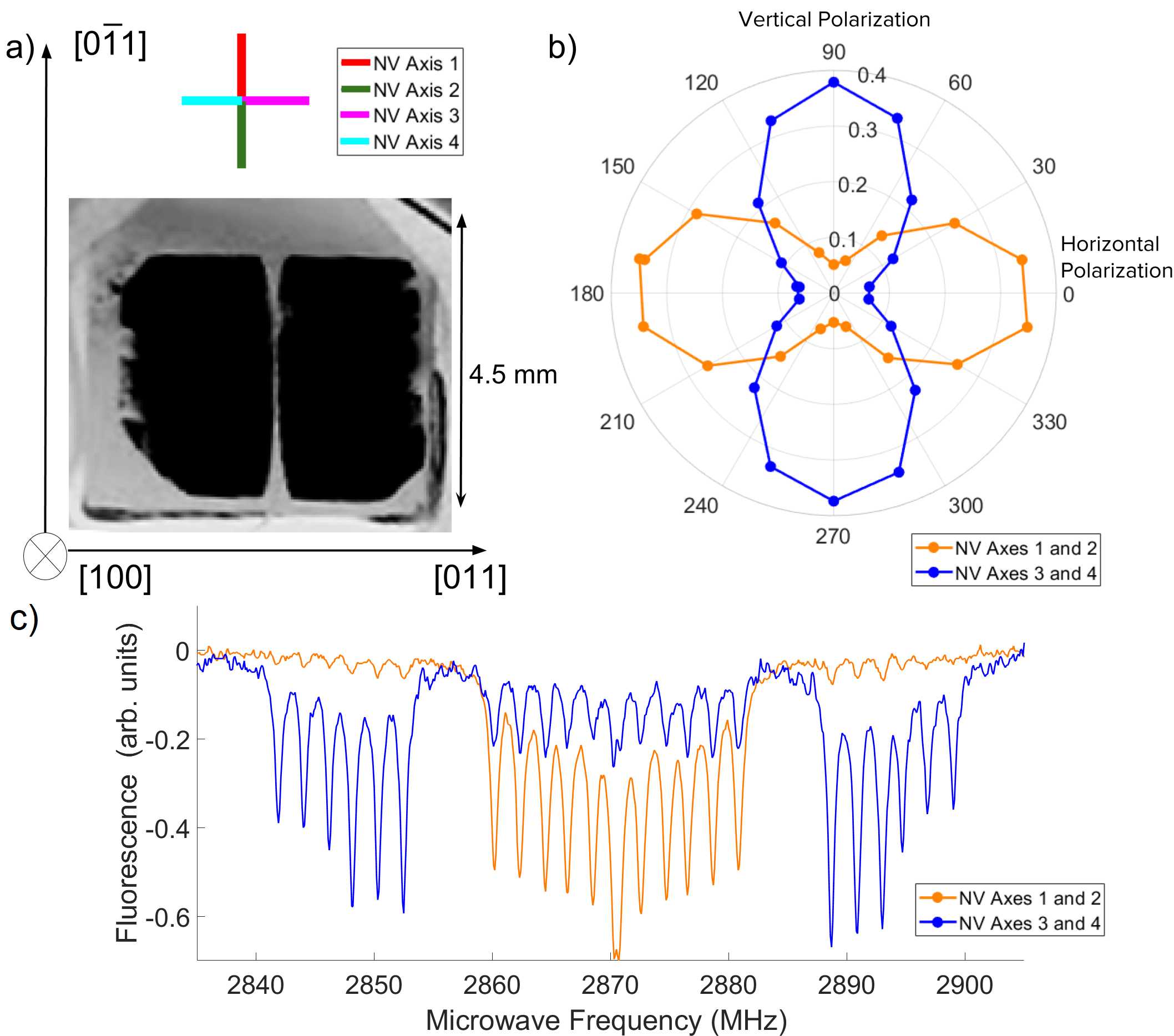}}
\caption{a)Four orientations of NV axes as seen looking down the [100] crytallographic axis and the electrodes printed on the (100) face of the diamond.  b)Magnitude of the optical coupling strength, measured as change in the ODMR contrast(arb. units) for the two pairs of the NV center orientations(directions labelled 1+2 and 3+4) as a function of the polarization of the exciting green laser. c) ODMR spectrum for horizontally (orange) and vertically (blue) polarized green excitation light. }
\label{fig:Electrodes}
\end{figure}

\section{Experimental Details}

Diamonds formed by chemical vapor deposition with a nominal nitrogen content of  $\approx$ 100 ppb were obtained from Element 6.\cite{E6} The diamonds are approximately 5x5x0.5 mm, cut so the 5x5 mm$^2$ faces are (100) planes and the 0.5x5 mm$^2$ edges are (011) planes, as shown in Fig. \ref{fig:Electrodes}. The diamonds were irradiated with an electron beam with an energy of 2 MeV and a fluence of 10$^{17}$ cm$^{-2}$ to generate the vacancies. They were then annealed for 2 hours at 875 K to encourage the coupling of vacancies with nitrogen nuclei. \cite{anneal} By integrating the visible absorption spectrum of the diamonds,\cite{conc} the density of NV states after processing was estimated to be 20 ppb. Two chromium electrodes, a few hundred nanometers in thickness, were photolithographically deposited on the top surface of one diamond. The electrodes are separated by a gap that is about 80 um wide at the center of the surface, growing to about  350 um near the edges. The patterned  electrodes create a slowly varying electric field in the diamond. %The electric field decreases with depth into the diamond, with the maximum electric field created at the surface.  

%\begin{figure}	
%{\includegraphics[scale = 0.3]{"24 resonances".png}}
%\caption{Twenty four resonances seen from the four possible NV orientations in the ensemble. Different NV orientations can be targeted by changing the polarization of the green laser. The numbered transitions from each NV axis correspond to the NV axes labelled in Fig. \ref{fig:Electrodes}}
%\label{fig:tfr}
%\end{figure}

 \begin{figure}
{\includegraphics[scale = 0.28]{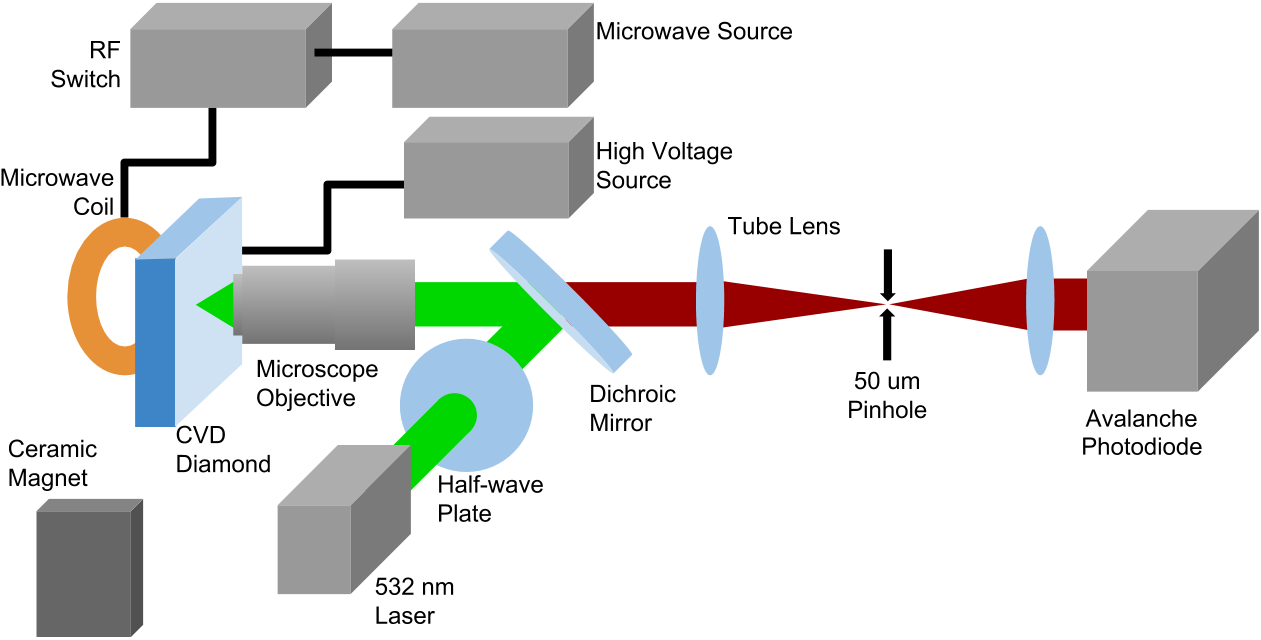}}
\caption{Experimental setup for electrometry using NV ensembles. The collected photoluminescence is spatially filtered to constrain the confocal volume being investigated. A high voltage source is used to apply a voltage of $\pm$ 3000 V across the electrodes. }
\label{fig:Schematic}
\end{figure}

\begin{figure}
{\includegraphics[scale = 0.32]{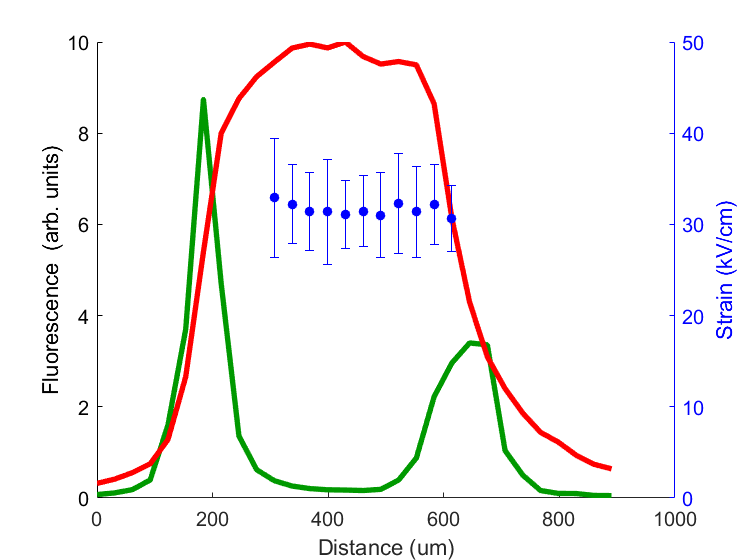}}
\caption{a) The green curve shows the magnitude of the laser light that is reflected from the top and bottom faces of the diamond. The red curves represent the intensity of the 637 nm ZPL. Curves are rescaled for clarity. The blue data points show the strain measured as a function of depth in the diamond. b) Difference in fluorescence at 0 V(blue), 2500 V (red) and their difference (yellow) }
\label{fig:StrainVsDepth}
\end{figure}

The experimental apparatus is shown schematically in Fig. \ref{fig:Schematic}. A specially constructed confocal microscope is used to investigate the ground state structure of NV ensembles. Green light from a frequency-doubled  Nd:YAG laser\cite{laserGlow} can be adjusted in power with a variable attenuator, while the angle of the linear polarization can be adjusted with a half-wave plate.  The light beam is brought onto the optical axis of a long working distance, 50x microscope objective by means of a dichroic mirror. The diamond is mounted in vacuum in a cryostat, with one electrode grounded to the cryostat as the potential of the other  is varied from 0 to  $\pm$ 3000~V DC. The rear, ($\bar{1}00$) face of the diamond is also grounded. The light is focused into the gap between the electrodes using the microscope objective located outside the cryostat.  Fluorescence is collimated by the objective and filtered through the dichroic beam combiner and by an optical longpass filter (Andover 590FG05, passing $ \lambda > 600$ nm).\cite{Andover} The light is spatially filtered with a 50 um pinhole before being sent to an avalanche photodiode\cite{Avalanche}. %Since diamond has a high dielectric constant, the generated electric fields have a large gradient. 

The microwave source is a digitally-synthesized oscillator (WindFreak Synth NV) \cite{SynthNV} amplified to 25 dBm\cite{Amplifier}. 
%The microwaves then pass through a digital switch, which enables digital modulation of the microwave amplitude at kHz rates. 
The microwaves are coupled to the diamond using a resonant loop of wire, about 1 cm from the diamond, located outside the cryostat. %that fits around the microscope objective, about 1 cm from the diamond, with the symmetry axis aligned with the optical axis. 

Since there are four possible crystallographic orientations of NV axes, we see four different sets of resonances from within the ensemble. The magnetic dipole transitions have the following selection rules:\cite{AcostaThesis} $\Delta m_s = \pm 1$ and $\Delta m_I = 0$; this leads to 6 ground state resonances. A total of 24 resonances (= 4 crystallographic orientations x 6 hyperfine transitions) can be observed in the ODMR spectrum.  Polarization of the green laser controls coupling strength\cite{Polarization} among the four NV orientations. This technique allows the selective excitation of NV centers, as shown in Fig \ref{fig:Electrodes} .

\begin{figure}
{\includegraphics[scale = 0.29]{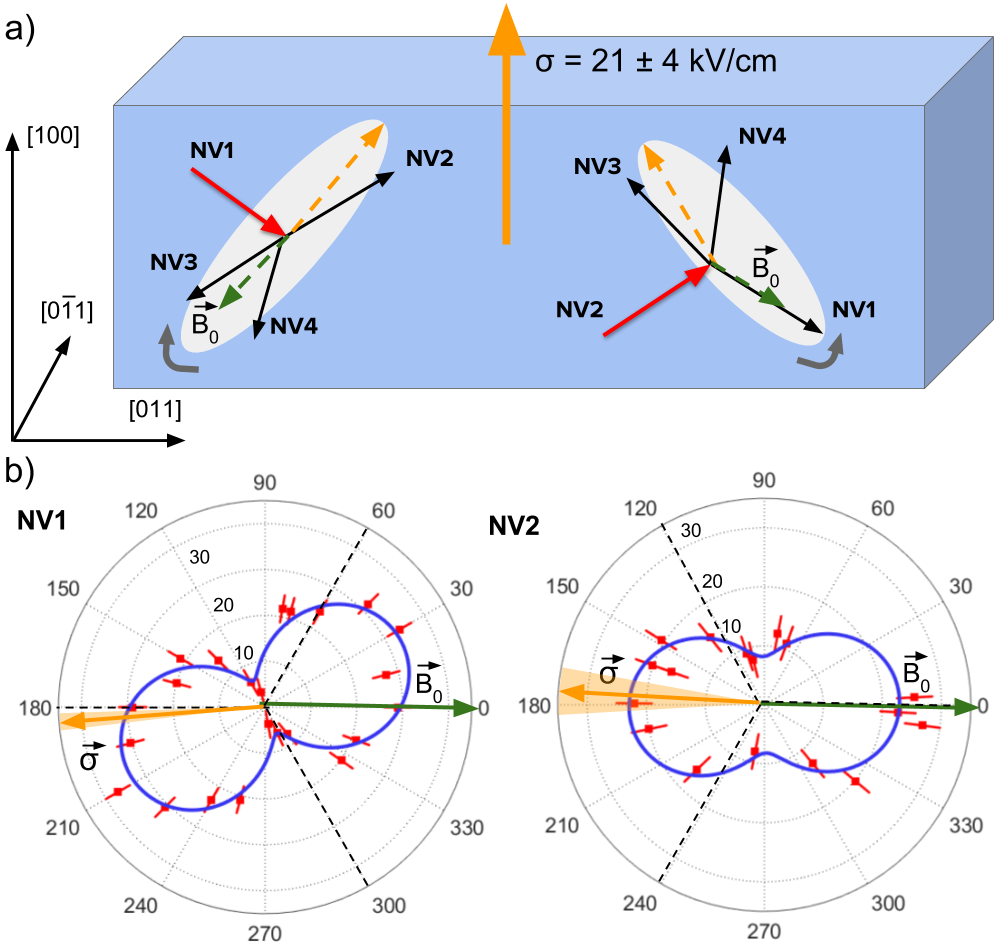}}
\caption{ a) Schematic illustrating the two NV orientations used to measure the strain direction. The white ellipses represent the plane in which the transverse magnetic field ($B_{\perp} \approx 1$ mT) is rotated. Strain direction as determined by the polar plots of the Stark shift is drawn in orange. The green arrows represent the initial direction of the magnetic field ($\theta = 0^{\circ}$ in the graphs below). b) Change in strain induced Stark shift as a function of the polar angle of the transverse magnetic field for the two NV orientations. Dashed black lines show the lines connecting the vacancy to the three adjacent carbon atoms. Here the strain induced Stark shifts are measured in kV/cm. The orange vector represents the strain as it projects on the transverse plane and the shaded region represents the uncertainty associated with the strain measurementw. Simulated data is shown as a the blue curve. }
\label{fig:StrainVsPolarB}
\end{figure}

%\begin{figure}
%{\includegraphics[scale = 0.33]{"Stark Shift".png}}
%\caption{Stark shift observed near a magnetic field of about $-77  \, \mu T$. The red(blue) spectrum corresponds to when the external electric field is turned on(off). The states are %labelled by their electronic and nuclear spin projections, $|m_s, m_I\rangle$. }
%\label{StarkShift}
%\end{figure}

%\begin{figure}
%{\includegraphics[scale = 0.53]{"Hyperfine shifts".png}}
%\caption{Stark shifts for six hyperfine resonances from a single NV axis obsereved from an ensemble in the middle of the electrodes. The shifts are larger for resonances that are near an avoided crossing. Here, the avoided crossing is near $B_z = 77 \,  \mu T$.}
%\label{StarkShiftsForSixHyperfine}
%\end{figure}

\section{Results}

The axial resolution is measured by moving the diamond through the focus of the microscope objective. As shown in Fig. 4a, the axial resolution is about 80 um. The transverse resolution, measured using a similar method, is about 1  um. % There is not a significant variation in the strain induced Stark shift through the sample. 
 %The charged and neutral centers can photoconvert. 
By measuring the fluoresence intensity at the NV$^-$ (637 nm) and NV$^0$ (575 nm) ZPL, we can measure changes in their relative concentrations. Closer to the surface, in regions of large electric field (E~$>$~40~kV/cm - see results below), there was marked decrease in the concentration of  NV$^0$ centers with a simultaneous increase in  NV$^-$ centers (up to 100\%). This effect was attenuated as we moved towards the grounded face of the diamond. We believe this effect is related to the charge state conversion observed in single NV centers because of field induced band-bending.\cite{ChargeStateConversion}

As shown in Ref. 5,  by solving the electronic Hamiltonian presented in Eq. (3), we can calculate the magnetic transition frequencies, $\omega_\pm$, between the states $m_s = 0$ and $m_s = \pm 1$

\begin{multline}
\hbar \omega_\pm  = D_{gs} + d_{gs}^\parallel \Pi_z \pm [ (\mu_B g_e B_z)^2 + (d_{gs}^\perp \Pi_{\perp})^2 \\ - \dfrac{(\mu_B g_e B_\perp)^2 }{D_{gs}} d_{gs}^\perp \Pi_{\perp} cos(2 \theta_B + \theta_\Pi) +  \dfrac{(\mu_B g_e B_\perp)^4 }{4D_{gs}^2}  ]^{1/2}
\end{multline}
Here, $\theta_\Pi$ and $\theta_B$ are the polar angles for the non-axial electric and magnetic fields, respectively. As shown below, we can use this dependence on the polar angle of the transerve magnetic and strain induced electric fields to estimate the direction of the strain field.   

%In our experimental setup for measuring external electric fields, the typical electric and magnetic fields are $E \approx 30$~kV/cm and $B_{\perp} \approx 200$ uT. In this case, we can use  $\dfrac{(\mu_B g_e B_\perp)^2 }{D_{gs}} <<  d_{gs}^\perp \Pi_{\perp}$ to simplify the above expression:

%\begin{multline}
%\hbar \omega_\pm  =  D_{gs} + d_{gs}^\parallel \Pi_z \pm [ (\mu_B g_e B_z)^2 + (d_{gs}^\perp \Pi_{\perp})^2 ]^{1/2}
%\end{multline}
%Note that in the above limit, the Stark shift is independent of the polar angle of the electric and magnetic fields. 

%This simplifies our analysis and we do not have to calculate the polar angles of the non axial electric and magnetic fields. This effect of the magnetic field orientation on the Stark shift  is discussed in other papers\cite{Dolde}.%

\subsection{Strain}
Since the total electric field measured in these experiments is the sum of the externally applied electric and effective electric field produced by the intrinsic strain, it is important to characterize the strain magnitude and direction for electric field measurements. As shown in Fig. 4a, the axial variation in the strain magnitude was measured by scanning the confocal microscope in the (100) crystallographic direction. 
%The collected fluorescence from the NV centers was also mapped during these measurements, this served as a rough measure for the NV concentration in the sample. No significant change  in fluorescence or strain was observed.
 
In order to measure the strain direction, the strain induced Stark shift was calculated for NVs along all four crystallographic directions. A largely orthogonal magnetic field with a small axial component of 77 uT was applied to isolate NV centers oriented along the four different (111) crystal directions. The magnitude of the strain was measured to be equal for all four axes. This narrowed down the strain heading to the three possible degenerate directions - (100), (010) and (001). Given this result, there is also the possibility of the strain to be isotropic. 

To complete our strain analysis, we used the dependence of the magnetic field polar angle, $\theta_B$ in eq.~(4) for a suitably large orthogonal magnetic field (1 mT), generated by a pair of Helmholtz coils and ceramic magnets. As the magnetic field was rotated in the non-axial plane, changes in the strain induced Stark shift were recorded. The polar plot from this experiment is shown in Fig. \ref{fig:StrainVsPolarB}. Fitting the graph to the energy shifts predicted by the Hamiltonian shows that the strain points in the (100) direction, out of the large face of the diamond. This direction is also parallel to the crystal growth direction. The total strain was determined to be $\sigma = 21 \pm 4$ kV/cm. 
%To minimize the inhomogenous broadening in the ODMR spectra, a low defect density diamond was used for this measurement. 

\subsection{Electric Field}

The ODMR transitions are shifted by the component of the electric field in the plane perpendicular to the NV axis. The Stark shift is typically small, a few Hz per kV/cm, but the shift is enhanced when the magnetic field is adjusted so that levels of the appropriate symmetry are nearly degenerate. As described in other work,\cite{Dolde} the shift is enhanced in the transitions to the $| m_S= +1,m_I = 0\rangle$ and $| m_S= -1,m_I = 0\rangle$ states when the axial magnetic field is zero. From the Hamiltonian given by Doherty,\cite{Doherty} we can also predict similar enhancements near $B_{z}     = 77$ uT where the $|-1, -1\rangle$ and $|+1,-1\rangle$ levels cross. Likewise, the $|-1, +1\rangle$ and $|+1,+1\rangle$  levels cross at  $B_{z}     = - 77 $ uT. Stark shift enhancement at the avoided crossings of these hyperfine sublevels is discussed in our previous work. \cite{Sharma}

\begin{figure}
{\includegraphics[scale = 0.32]{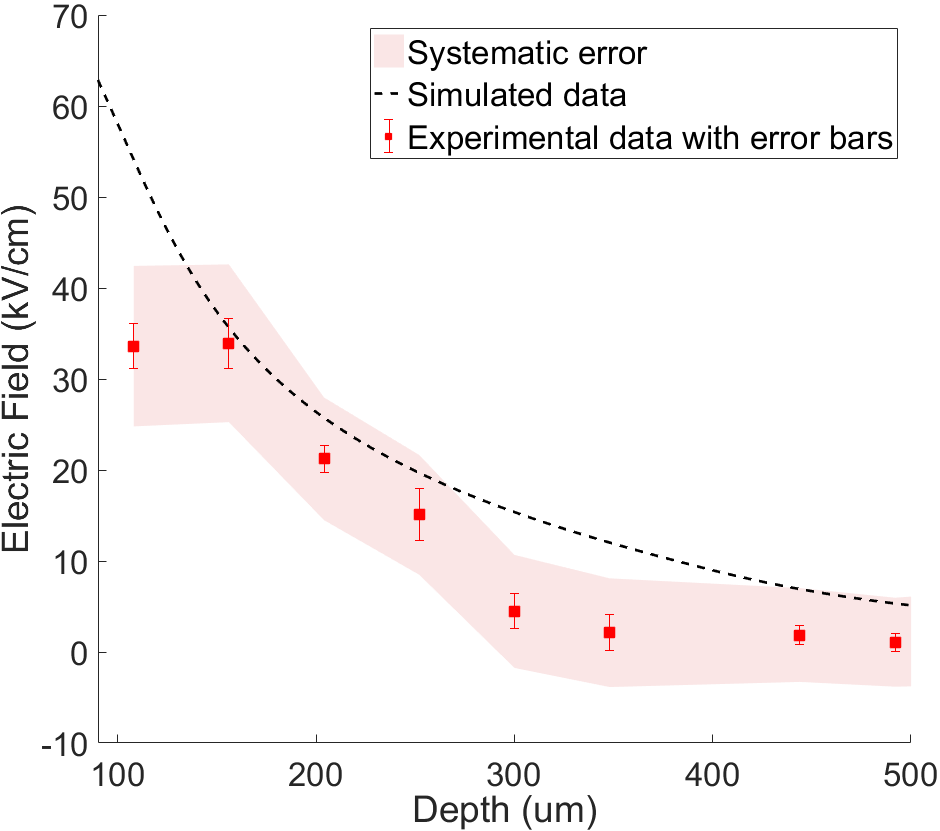}}
\caption{Electric field measured as a function of depth the diamond. The simulated data is obtained from finite element analysis done in COMSOL.\cite{COMSOL} }
\label{ElectricFielDepth}
\end{figure}

The electrodes were patterned to produce variation in the electric field. A map of the electric field was produced by moving the focal spot axially into the diamond and measuring the Stark shift for the same voltage difference across the electrodes, as shown in Fig.\ref{ElectricFielDepth}. %The resolution of the image produced is dictated by the numerical aperture of the objective and the size of the pinhole used. 
The electric field measurements were obtained by subtracting the strain($\vec{\sigma}$) induced Stark shift from the total Stark shift($\vec{\Pi} = \vec{E} + \vec{\sigma}$).

 The largest contributor to the systematic error was the uncertainty in the electric dipole moment, $d_{\perp}$~=~$17 \pm$~$ 3$~Hz~cm/V. With just statistical errors, the typical uncertainty in the electric field measurement was $\sigma \, = \, $ 2~kV/cm. After including systematic uncertainties, the cumulative error was 5 ~kV/cm. This shows that we can map electric fields throughout the whole volume of the diamond. 
%The total internal field seen by the NV centers is the vector sum of a contribution from the field generated by the voltage applied across the electrodes and the apparent electric field produced by internal strain in the diamond.  The internal electric field within the diamond is reduced compared to the field applied to the diamond by the diamond polarizability. The fields reported here are internal electric fields, related to the external field by $E_{int} = E_{ext}/\kappa$, where $\kappa \approx 5.7$ is the relative permittivity of diamond. 
%In order to measure the effect of the applied electric field, measurements are taken at a pair of voltages, typically 0 and 2000 V. The frequencies of the six hyperfine components of a particular NV orientation, measured at at least two applied voltages, are used to determine $D_0, B_{||}, E_{\perp}$ and strain component $S_\perp$. 

%The sensitivity of the innermost splitting can be found by evaluating the Hamiltonian at a number of electric fields, or, far from the avoided crossing, by taking the derivative of Eq. 5 with respect to the electric field. At each avoided crossing, the splitting has a maximum value of $34 \, Hz/(V/cm)$. For an uncertainty in the splitting of about 5 kHz and perfect knowledge of the magnetic field, the corresponding uncertainty in the electric field is $147 \, \, V/cm$ . Far from the crossings, the sensitivity scales as $E^\perp / B_z$ , with a numeric value of $0.68 \,   Hz/(V/cm)$ at $E^\perp = 30 \, kV/cm$ and $B_z = 1 \, mT$ (and $B_x = B_y = 0$).
%
Future work will show how this technique can use NV center ensembles to serve as vector electrometers. The orthogonal component of the electric field from each of the four NV axes can be used to recreate the electric field. We also plan to investigate the influence of electric and magnetic fields on decoherence rates of the NV center.
%The magnitude of level repulsion in the ground state resonances can also be used as a measure of the inherent strain in the diamond. 

\section{Acknowledgements}

This material is based upon work supported by the Department of Energy under Award Number  DE-SC0011266.\cite{Disclaimer} This work is also in supported in part by NSF PHY-1506416 (Sharma, S. and Beck, D.H.). This work was carried out in part, at the Frederick Seitz Material Research Laboratory Central Research Facilities, University of Illinois. 

\pagebreak


\begin{thebibliography}{9}
\bibitem{Disclaimer}  Disclaimer: This report was prepared as an account of work sponsored by an agency of the United States Government.  Neither the United States Government nor any agency thereof, nor any of their employees, makes any warranty, express or implied, or assumes any legal liability or responsibility for the accuracy, completeness, or usefulness of any information, apparatus, product, or process disclosed, or represents that its use would not infringe privately owned rights.  Reference herein to any specific commercial product, process, or service by trade name, trademark, manufacturer, or otherwise does not necessarily constitute or imply its endorsement, recommendation, or favoring by the United States Government or any agency thereof.  The views and opinions of authors expressed herein do not necessarily state or reflect those of the United States Government or any agency thereof.

			

%
\
\end{thebibliography}
\end{document}